\documentclass[preprint]{elsarticle}

\usepackage[utf8]{inputenc}  
\usepackage[T1]{fontenc}     
\usepackage{lmodern}         
\usepackage{microtype}       
\usepackage[scaled]{helvet} 
\usepackage[english]{babel}  
\usepackage{graphicx}        
\usepackage[bookmarks=false]{hyperref}            
\usepackage{comment}         
\biboptions{sort&compress}   

\journal{TBD}

\begin{document}

\begin{frontmatter}

\title{Stake Your Claim: Zero-Trust Validator Deployment Leveraging NFTs and Smart Contracts in Proof-of-Stake Networks}

\author{Scott Seidenberger\fnref{label1,label2}%
}
\author{Alec Sokol\fnref{label1}%
}
\author{Anindya Maiti\fnref{label2}%
}
\fntext[label1]{186Kapital Group, Washington, DC%
}
\fntext[label2]{University of Oklahoma, Norman, OK%
}
\begin{abstract}
We present a novel method for a multi-party, zero-trust validator infrastructure
deployment arrangement via smart contracts to secure Proof-of-Stake (PoS)
blockchains. The proposed arrangement architecture employs a combination of
non-fungible tokens (NFTs), a treasury contract, and validator smart contract
wallets to facilitate trustless participation in staking mechanisms. The NFT
minting process allows depositors to exchange their capital for an NFT representing
their stake in a validator, while the treasury contract manages the registry of
NFT holders and handles rewards distribution. Validator smart contract wallets
are employed to create a trustless connection between the validator operator and
the treasury, enabling autonomous staking and unstaking processes based on
predefined conditions. In addition, the proposed system incorporates protection
mechanisms for depositors, such as triggered exits in case of non-payment of
rewards and a penalty payout from the validator operator. The arrangement
benefits from the extensibility and interoperability of web3 technologies, with
potential applications in the broader digital ecosystem. This zero-trust staking
mechanism aims to serve users who desire increased privacy, trust, and flexibility
in managing their digital wealth, while promoting greater decentralization and
transparency in the PoS ecosystem.
\end{abstract}
\end{frontmatter}


\section{Introduction} 

The collection of protocols, networks, and applications that define the emerging web3 technology stack \cite{voshmgir2020token} has brought about a paradigm shift in the way users interact with digital platforms and services. Decentralization, privacy, and security have become key features at the forefront of restoring the original tenants of the web. Underpinning the functioning of the web, and economic interactions in general, is trust. Therefore, the creation and widespread adoption of trustless systems offers significant benefits to all participant stakeholders. 

Distributed ledgers and smart contracts are two critical components of web3 that enable the creation of trustless automation, minimizing or even eliminating counterparty risk in various types of transactions. By decentralizing the validation and verification processes of transactions \cite{almakhour2020verification}, distributed ledgers provide a transparent and tamper-resistant infrastructure where multiple parties can safely interact without having to rely on a central authority. This decentralized approach enhances the resiliency and security of the system while mitigating the risk of single points of failure.

Smart contracts, as self-executing agreements with contract terms implemented as code, further facilitate trustless automation. These programmable contracts can be made immutable once deployed and execute autonomously when called. When architected as part of a larger system, smart contracts can minimize or eliminate the need for human intermediaries while increasing transparency and observability of the system as a whole. As a result, smart contracts bring about increased efficiency, reduced costs, and enhanced trust among parties engaging in transactions.

Together, distributed ledgers and smart contracts lay the foundation for a more transparent, and secure digital landscape. By automating trust and minimizing counterparty risks, they empower individuals and organizations to engage in various economic activities with increased confidence. 

Some of the largest and most widely adopted distributed ledgers supporting smart contracts are secured by Proof-of-Stake (PoS) consensus \cite{king2012ppcoin} mechanisms. In PoS networks, validators are instrumental in securing and maintaining the system's integrity. They stake their tokens in exchange for the right to earn rewards from producing valid blocks. However, traditional staking solutions often impose substantial requirements in terms of upfront capital, technical expertise, and custodial arrangements, thus creating barriers for most users seeking to participate in securing the network. To address these challenges and make PoS networks more accessible, innovative approaches to validator infrastructure deployment are essential. 


\section{Staking Mechanisms} 

There are three main schemas to deploy, stake, and operate validators: 

\textbf{Solo Staking.} In solo staking, a single party deploys their own infrastructure and directly interacts with a network in order to run their validator. The individual is responsible for operation, configuration management, and security, as well as providing all upfront capital required. For the security of the decentralized network, this option is the most robust. Solo stakers that run validators on their own hardware are more likely to increase the geographic diversity of nodes, with different power sources, internet connections, configurations, and political jurisdictions. 

\textbf{Validator-as-a-Service (VaaS).} In this category, a two-party system is formed where the capital is provided by the first party, who then delegates the operation of the validator to a second party. This delegation can either be native to the protocol (Delegated Proof-of-Stake DPoS) or can be arranged via a managed service. Additionally, arrangements can be classified as either custodial or non-custodial. In both custodial arrangements, the validator operator has control of specific signing keys that allow them to operate on behalf of the depositor and their stake. The defining feature of non-custodial is the capital depositor retains custody of the withdrawal keys needed to unstake the validator and regain control of their original stake. 
    
In this schema, trust is introduced around the handling of validator operation. The trust introduces risk that the VaaS provider may act dishonestly, or act in a risky manner that puts the capital at risk of being slashed or rewards being lost. VaaS providers may also not have their code or operating model open source or audited, which leads to further trust. 

\textbf{Staking Pools.} In response to substantial capital requirements (whether that be through upfront stake or infrastructure maintenance), staking pools offer an arrangement where a multi-party system is formed to pool capital to launch a validator. The specifics of this type of arrangement vary significantly depending on the construction of the pool and economic design decisions. Some may be centralized, most often run by large cryptocurrency exchanges that take custody of users' capital in order to run validators in exchange for a variable yield. Other pools are governed by distributed autonomous organizations (DAOs) that use smart contracts to track user deposits and their share of rewards in order to mediate the pooling of capital. 
    
Additionally, many large pools offer a \emph{liquid staking derivative}, or a token that represents a deposited amount of capital into the pool. These derivative tokens allows users to have some degree of liquidity over their deposited assets, but can carry their own additional inherent risk. This multi-party arrangement relies on trusting the successful orchestration of the complex social, technical, and economic dimensions. 



\section{Proposed Architecture} 

\textbf{Overview.} We propose a novel method for a multi-party, zero-trust validator infrastructure deployment arrangement via smart contracts to secure Proof-of-Stake blockchains. This arrangement fills a gap in staking options, sharing characteristics with both VaaS arrangements and staking pools. The system is implemented by a series of interlocking, unmodifiable smart contracts. 

\textbf{Network Prerequisites.} In order to implement the architecture, the network must allow: 

\begin{itemize}
    \item A turing complete virtual machine.
    \item Smart contracts to invoke other smart contracts.
    \item Designated smart contract methods to be restricted to NFT holders.
    \item Smart contracts to perform the staking action for a validator.
    \item A validator's withdrawal address to be immutable.
    \item A validator's withdrawal address to be a smart contract.
    \item Smart contracts to have a sense of time (eg. via block height or provided by oracle).
\end{itemize} 

\textbf{NFT Mint.} The capital to be staked is collected via a non-fungible token (NFT) minting contract which transfers depositors' capital to a treasury smart contract in exchange for an NFT with metadata associated with the treasury. The holder of the NFT has effective control over both the initially deposited capital as well as the rewards from the validator. An NFT allows for unique and indivisible tokens representing fractional ownership in the validator, carrying attributes that carry more data and enable extensibility and interoperability in the broader ecosystem. 

\textbf{The Treasury.} The Treasury is responsible for handling the registry of NFT owners, and contains the methods for disbursing rewards, staking, and unstaking validators. Rewards earned by the validator automatically flow to the Treasury contract, which are then claimable by the NFT holder. The Treasury enables the specifics of the staking arrangement to be constructed prior to the mint, and once constructed will be immutable and transparent to those participating in the mint. 

\textbf{The Validator Smart Contract Wallet.} To stake and unstake a validator, the Treasury calls a validator smart contract, which uses account abstraction to hold funds and self-execute code on the blockchain’s virtual machine. The core of the validator smart contract wallet is that the they can initiate transactions themselves on-chain, serving as a trustless intermediary between the validator operator and the treasury. 

In this arrangement, depositors do not have to create their own keys, they are created and managed by a smart contract wallet on their behalf. This is trustless because the partial owners of the same validator do not have to trust the others in the pool nor the validator operator who is assigned signing keys from the validator smart contract. This allows for truly zero-trust, partial, staking through direct validator ownership. 

\textbf{Depositor Protection via Triggered Exit.} In the event that the treasury does not receive the rewards expected from the validators, the validator smart contract wallet can autonomously unstake itself and return principal as well as accrued rewards to the NFT holders via the treasury. The extensibility of this arrangement allows for more logic and economic incentives to be added to the validator smart contract wallet, such as enabling an insurance payment from an escrow account to be paid out to NFT holders for lack of validator performance. The autonomous unstaking process maximally protects depositors while also aligning incentives for validator operators. 



\section{Specification} 

\begin{figure}[h]
    \noindent\makebox[\textwidth]{%
    \includegraphics[width=\textwidth]{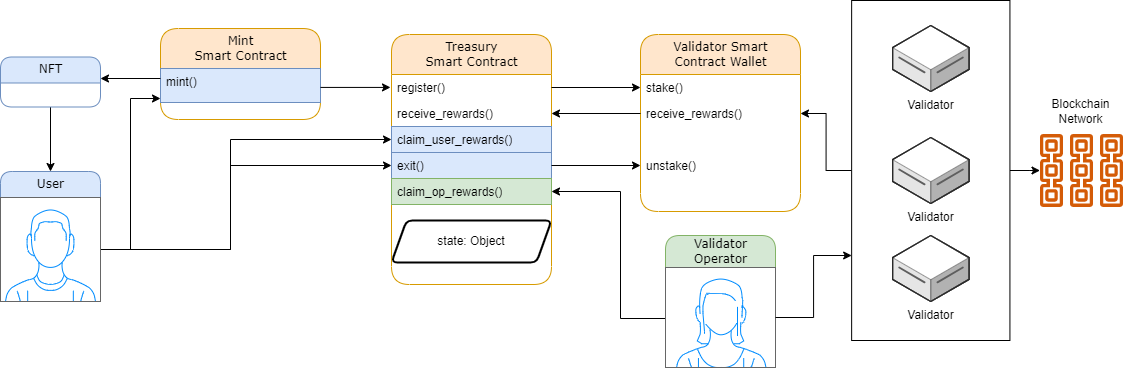}}
    \caption{Architecture Diagram.}
  \end{figure}

$m$: Number of validators assigned to the same treasury.

$R_j$: Total rewards earned by the j-th validator.

$F$: Validator operator's reward fee ratio in range (0,1). 

The total operator's fee revenue ($S$) can be calculated as:

$$
S = \sum_{j=1}^{m}{R_j} * F
$$ 

$n$: Number of NFTs.

$C_i$: Capital contributed by the i-th NFT holder. 

The share of rewards for the i-th NFT holder ($r_i$) can be calculated as:

$$
r_i = \frac{C_i}{\sum_{i=1}^{n}C_i} \times (\sum_{j=1}^{m}{R_j} \times (1- F))
$$ 



\section{Risks} 

\textbf{Smart Contract Security.} Any implementation of this arrangement should adhere to smart contract security best practices. This includes code modularity, proper access controls, and safe arithmetic operations. It is advised that depending on the network chosen to deploy such smart contracts, use of NFT standards and minting contracts be used. This risk can be mitigated by soliciting thorough audits from reputable third-party auditors. 

\textbf{Validator Performance.} Depending on the network, poorly performing or misbehaving validators may risk capital via slashing. The zero-trust nature of this system ensures that the conditions of the arrangement are self-executing, but do not inherently completely prevent slashing risk. The validator operator should show evidence that they have sufficient safeguards in place to prevent slashing. Additional economic alignment can be accomplished by having validator operators post a certain collateral amount in escrow to the Treasury to disincentive poor performance and provide some slashing protection. 

An open research area is having decentralized oracle networks verify validator performance metrics, providing them on-chain for use within this system.



\section{Extensibility} 

\textbf{Target Audience.} The target audience for this arrangement are those who desire a zero-trust and permissionless staking mechanism. Since the ownership stake in the validator is encoded in a dynamic, utility NFT, it gives the holder composability and interoperability within the emergent digital ecosystem. The composability aspect allows others to build on top of this infrastructure and what it represents on-chain, unlocking new digitally native value propositions. This type of solution is also aimed at those who want added layers of privacy and trust when building their digital wealth. For example, zero-trust architecture is much more valuable where trust in current institutions and financial networks are strained. 

\textbf{Composability.} The proposed architecture, which utilizes NFTs to represent trustless ownership of validators on PoS blockchains, offers several advantages in terms of composability and interoperability. These characteristics are essential in the rapidly evolving web3 landscape, where decentralized applications (dApps), smart contracts, NFTs, and the metaverse are continuously interacting and building on one another.

NFT-based validator stakes can be utilized as collateral in decentralized finance (DeFi) applications, allowing users to borrow, lend, or participate in various financial services while maintaining their stake in the validator. This enables users to unlock additional value from their staked assets without having to unstake or sell their positions. 

Moreover, the NFT-based validator stakes can be integrated into DAOs as a governance mechanism, giving stakeholders voting rights proportional to their ownership stakes in validators. This would enable differentiated stakeholder decision-making processes and encourage active participation in the governance of the underlying blockchain network. DAOs could also collectively invest in a validator NFT to build digitally native wealth for digitally native organizations.  

\textbf{Interoperability.} By encoding validator stakes as NFTs, the proposed arrangement facilitates interoperability between various web3 technologies. One potential area of is the integration of validator NFTs within virtual worlds and metaverse platforms. Users could showcase their validator stakes in digital galleries, trade or sell them in virtual marketplaces, or even use them as in-game assets with unique utility. This not only adds another layer of functionality to the validator NFTs but also fosters a more immersive and engaging experience for users in the metaverse.  

Furthermore, the NFT-based validator stakes can be utilized across different blockchain networks through cross-chain bridges and atomic swaps, allowing users to easily transfer and manage their assets across various ecosystems. This enhances the overall user experience, reduces friction in asset management, and fosters greater collaboration between different blockchain networks.

By leveraging the unique capabilities of NFTs and smart contracts, this system has the potential to drive greater adoption of PoS networks, democratize access to staking opportunities, and contribute to the development of a more decentralized, equitable, and interconnected digital landscape. 



\section{Conclusion} 

The proposed arrangement architecture presents a novel, multi-party, zero-trust validator infrastructure deployment for securing Proof-of-Stake blockchains. By leveraging the unique characteristics of NFTs and smart contracts, this system offers a more accessible, flexible, and secure staking solution compared to traditional approaches. In addition, the arrangement's extensibility and interoperability make it attractive to a wide range of users, from those seeking privacy and security to developers building innovative solutions on top of the infrastructure.

The growth and adoption of web3 technologies have the potential to bring about a more transparent and secure digital landscape. By providing novel staking solutions such as this, we can address current gaps and grow the web3 ecosystem. As the space continues to evolve, it is crucial to keep exploring solutions that grow the decentralization and strength of the networks that will power the web of tomorrow.


\bibliographystyle{ACM-Reference-Format}
\bibliography{main}

\end{document}